\documentclass[conference]{IEEEtran}\IEEEoverridecommandlockouts
\usepackage{etoolbox}
\usepackage{algorithmic}
\usepackage{algorithm}

\makeatletter
\patchcmd{\@makecaption}
  {\scshape}
  {}
  {}
  {}
\makeatletter
\patchcmd{\@makecaption}
  {\\}
  {.\ }
  {}
  {}
\makeatother
 \usepackage{amsmath,amssymb}
 \usepackage{subfigure}
 \usepackage{graphicx,graphics,color,psfrag}
 \usepackage{cite,balance}
 \usepackage{caption}
 \allowdisplaybreaks
 \usepackage{algorithm}
 \usepackage{accents}
 \usepackage{amsthm}
 \usepackage{bm}
 \usepackage{algorithmic}
 \usepackage[english]{babel}
 \usepackage{multirow}
 \usepackage{enumerate}

 \usepackage[T1]{fontenc}
 \usepackage{aecompl}

 \usepackage{cases}
 \usepackage{bm}
 \usepackage{stfloats}
 \usepackage{dsfont}
 \usepackage{color,soul}
 \usepackage{amsfonts}
 \usepackage{cite,graphicx,amsmath,amssymb}
 \usepackage{subfigure}
 \usepackage{fancyhdr}
 \usepackage{hhline}
 \usepackage{graphicx,graphics}
 \usepackage{array,color}
 \usepackage{amsmath}
\usepackage{float}
\usepackage{amssymb}
\usepackage{amsmath}
\usepackage{amsthm}
\usepackage{amsfonts}
\usepackage{graphicx}

\usepackage{epstopdf}
\usepackage{cite}
\usepackage{amsmath,bm}
\usepackage{subfigure}
\usepackage{graphicx}
\usepackage{color}
\usepackage{graphicx}
\usepackage{calc}
\usepackage{caption}

\begin{document}

\title{MambaJSCC: Deep Joint Source-Channel Coding with Visual State Space Model}
\author{Tong Wu$^{*}$, Zhiyong Chen$^{*}$, Meixia Tao$^{*}$, Xiaodong Xu$^{\dag}$, Wenjun Zhang$^{*}$, and Ping Zhang$^{\dag}$\\
		$^{*}$Cooperative Medianet Innovation Center, Shanghai Jiao Tong University, Shanghai, China\\
$\dag$Beijing University of Posts and Telecommunications, Beijing, China\\   
		Email: \{wu\textunderscore tong, zhiyongchen, mxtao, zhangwenjun\}@sjtu.edu.cn, xuxd@pcl.ac.cn, pzhang@bupt.edu.cn}
\maketitle
\begin{abstract}
Lightweight and efficient deep joint source-channel coding (JSCC) is a key technology for semantic communications. In this paper, we design a novel JSCC scheme named MambaJSCC, which utilizes a visual state space model with channel adaptation (VSSM-CA) block as its backbone for transmitting images over wireless channels. The VSSM-CA block utilizes VSSM to integrate two-dimensional images with the state space, enabling feature extraction and encoding processes to operate with linear complexity. It also incorporates channel state information (CSI) via a newly proposed CSI embedding method. This method deploys a shared CSI encoding module within both the encoder and decoder to encode and inject the CSI into each VSSM-CA block, improving the adaptability of a single model to varying channel conditions. Experimental results show that MambaJSCC not only outperforms Swin Transformer based JSCC (SwinJSCC) but also significantly reduces parameter size, computational overhead, and inference delay (ID). For example, with employing an equal number of the VSSM-CA blocks and the Swin Transformer blocks, MambaJSCC achieves a 0.48 dB gain in peak-signal-to-noise ratio (PSNR) over SwinJSCC while requiring only 53.3\% multiply-accumulate operations, 53.8\% of the parameters, and 44.9\% of ID.
\end{abstract}

\section{Introduction}
Deep learning-based joint source-channel coding (JSCC), which combines source and channel coding into a single process using neural networks to optimize end-to-end transmission over wireless channels, has recently gained significant attention as a means to increase the efficiency of information transmission. With the aid of deep learning, JSCC can effectively extract and utilize the semantic features of data, which are critical for understanding the meaning of information based on specific task requirements. Therefore, JSCC is emerging as a key technology for enabling semantic communications \cite{Qin2,Qin1}, a pivotal element in the development of sixth-generation (6G) wireless networks\cite{shao,wei,CDDM}.

Most of the recent research on JSCC focuses on finding suitable schemes to maximize end-to-end performance. For text transmission, \cite{text} proposes a JSCC framework with a joint iterative decoder to enhance text transmission quality. For video transmission, \cite{video} develops a deep video semantic transmission system based on JSCC to maximize overall performance. For image transmission, DeepJSCC, proposed in \cite{gundu2019}, is based on convolutional neural networks (CNNs) and outperforms traditional separation-based schemes, e.g., JPEG2000 and capacity achieving channel code. To further enhance the performance of JSCC, SwinJSCC is proposed in \cite{SwinJSCC}, which replaces CNNs with Swin Transformers\cite{Ze}.

Designing deep learning-based JSCC entails considerations beyond end-to-end performance, and it also necessitates evaluating the complexity of artificial intelligence (AI) algorithms and their adaptability to the dynamic changes of wireless channels. The performance of DeepJSCC is considered unsatisfactory due to the limited capacity of CNNs. In contrast, SwinJSCC achieves better performance but incurs higher computational overhead and has a larger model parameter size. To overcome performance degradation caused by dynamic channel conditions, attention modules are introduced into the JSCC architecture in \cite{Xu,SwinJSCC}, enabling it to incorporate channel state information (CSI) for adapting to changes in the channel signal-to-noise ratio (SNR). However, these attention modules introduce a significant number of additional parameters and computational overhead, greatly increasing the complexity of a single model.

To this end, this paper proposes a novel lightweight and efficient deep JSCC scheme, named MambaJSCC, using the visual state space model (VSSM) \cite{VSSM} and CSI embedding for image transmission over wireless channels. Recently, Mamba is proposed in \cite{S6} based on state space models (SSMs), where the state space is adjusted to a selective structure state space, enabling the models to focus on relevant information in an input-dependent manner. Mamba achieves remarkable performance in natural language processing tasks with linear complexity. Motivated by Mamba, we design a VSSM with channel adaptation (VSSM-CA) block as the backbone for MambaJSCC. Specifically, VSSM-CA integrates 2D images with state space to capture their global information. Meanwhile, a novel CSI embedding method is proposed for VSSM-CA, which injects CSI into the VSSM-CA block in a low-complexity manner, allowing it to adapt to changes in wireless channels. Leveraging the powerful VSSM-CA backbone, we design MambaJSCC, which features a hierarchical structure with varying numbers of patches of different sizes to effectively encode and decode images. We also develop a 2D patch division module in MambaJSCC that reverses the process of patch merging in the encoder to reconstruct the source image. Experimental results demonstrate that the proposed MambaJSCC outperforms SwinJSCC, and it has significantly lower parameter size, computational overhead, and inference latency (ID) compared to SwinJSCC.
\begin{figure*}[t]
  \begin{center}
    \includegraphics[width=16cm,height=8.5cm]{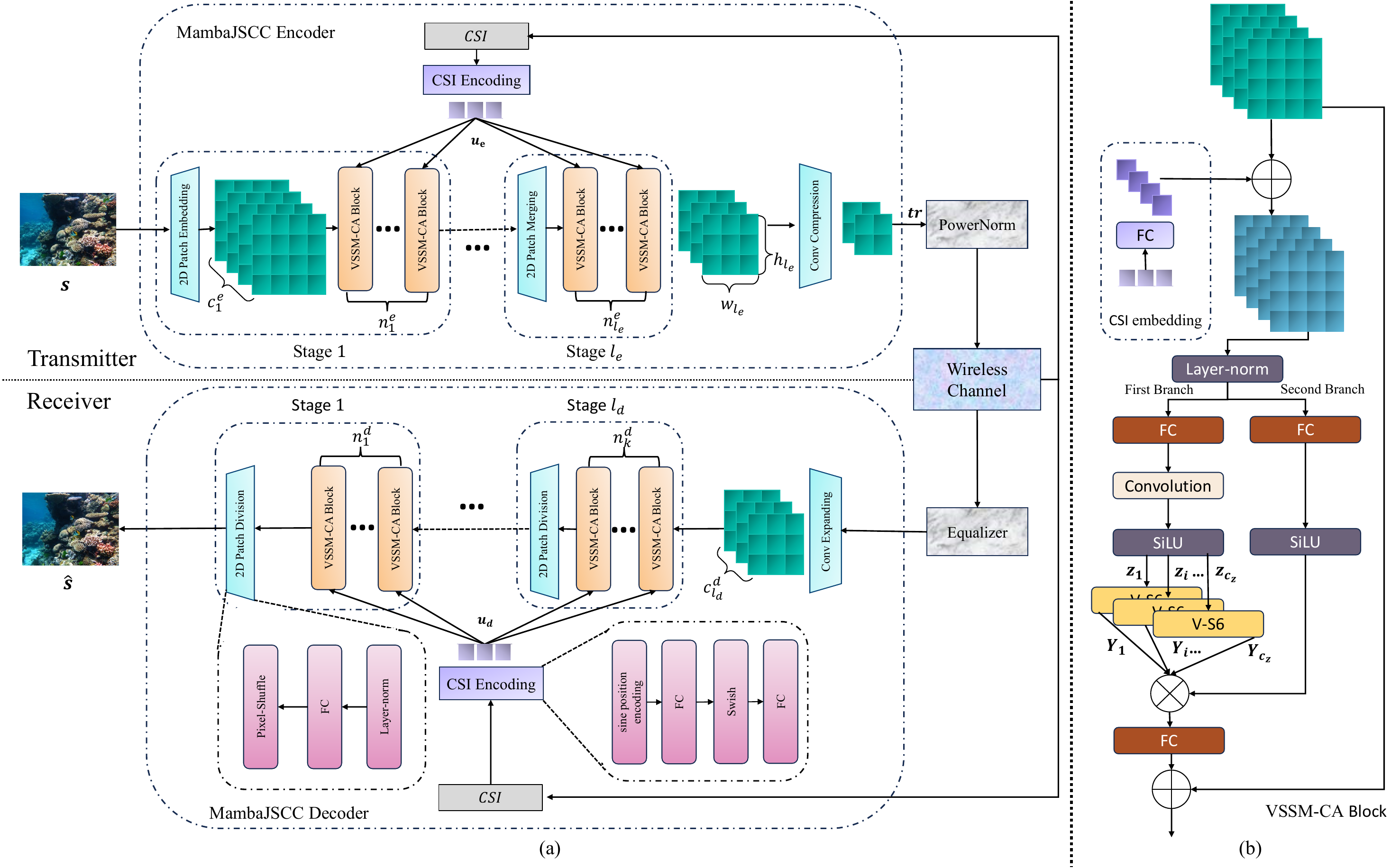}
  \end{center}
    \caption{{(a) The overall architecture of the proposed MambaJSCC. (b) The structure of the VSSM-CA block.}}
    \label{MambaJSCC}
    \vspace{-0.5 cm}
\end{figure*}

\section{System Model of MambaJSCC}
\subsection{State Space Model}
State space models are a class of sequential neural network models inspired by linear time-invariant (LTI) systems and are widely related to CNNs and recurrent neural networks (RNNs). Generally, LTI systems map a one-dimension function or sequence $x(t)\in\mathbb{R}$ to $y(t)\in\mathbb{R}$ through a latent state space ${h}(t)\in\mathbb{R}^N$. Here, $N$ is the dimension of the latent state space. The process can be formulated using linear ordinary differential equations (ODEs) as follows:
\begin{align}\label{LTI}
  {h^{\prime}}(t)=\mathbf{A}h(t)+\mathbf{B}x(t),~~ y(t)=\mathbf{C}h(t)+\mathbf{D}x(t),
\end{align} 
where $\mathbf{A}\in \mathbb{R}^{N\times N}$, $\mathbf{B}\in \mathbb{R}^{N\times 1}$, $\mathbf{C}\in \mathbb{R}^{1\times N}$, and $\mathbf{D}\in \mathbb{R}^{1\times 1}$ are the state matrix, input matrix, output matrix, and feed-forward matrix, respectively.

The OEDs represent continuous-time systems that require a discretization process to be integrated into deep learning frameworks. A widely-adopted discretization method is the zero-order hold (ZOH) rule, which utilizes a timescale parameter $\mathbf{\Delta} \in \mathbb{R}^{N \times N}$ to discretize the continuous parameters $\mathbf{A}$ and $\mathbf{B}$ into their discrete parameters $\mathbf{\bar{A}}$ and $\mathbf{\bar{B}}$. The process can be formulated as follows:
\begin{align}
  \mathbf{\bar{A}}=\exp(\mathbf{\Delta}\mathbf{A}),~~
 \mathbf{\bar{B}}=(\mathbf{\Delta}\mathbf{A})^{-1}(\exp(\Delta\mathbf{A})-\mathbf{I})\mathbf{\Delta}\mathbf{B}.
\end{align}
Then, (\ref{LTI}) can be rewritten in discrete-time form as follows:
\begin{align}\label{D-LTI}
  h_{t}=\mathbf{\bar{A}}h_{t-1}+\mathbf{\bar{B}}x_t,~~
 y_t=\mathbf{C}h_t+\mathbf{D}x_t.
\end{align}

\subsection{An Overview of the Proposed MambaJSCC}
An overall architecture of MambaJSCC is illustrated in Fig. \ref{MambaJSCC}(a). In the transmitter, the input image $\mathbf{s}\in\mathbb{R}^{3 \times H \times W}$ is encoded into the channel input signal $\mathbf{tr}\in\mathbb{C}^{c \times h \times w}$ by the MambaJSCC encoder, where $H$ and $W$ represent the height and width of the source image, respectively. This encoding process involves one CSI encoding module, $l_e$ coding stages, and a convolution compression module. The CSI encoding module encodes the CSI into a CSI vector $\mathbf{u_e}\in \mathbb{R}^{m}$, where $m$ is the vector length. The vector is then fed into all the VSSM-CA blocks to adapt to channel conditions. 

In the first coding stage, the input image $\mathbf{s}$ is initially fed into a 2D patch embedding module, which outputs $c_1^e$ patches, each of size $h_1\times w_1$. Subsequently, $n_1^e$ VSSM-CA blocks, which incorporate the CSI vector $\mathbf{u_e}$, are employed to process these patches. These blocks adaptively extract and encode the features from the patches into new patches based on the CSI, with the number and size of the patches remaining unchanged. The output patches are then fed into Stage 2, where the first module is the patch merging module. This module merges the $c_1^e$ patches of size $h_1 \times w_1$ into a greater number of patches $c_2^e$, each of a smaller size $h_2 \times w_2$, aiming to capture more refined information. Following this, $n_2^e$ VSSM-CA blocks, with the additional input of the CSI vector $\mathbf{u_e}$, are employed on these patches. This stage is repeated ($l_e-1$) times in the MambaJSCC encoder as Stage $k$, $k=2,3,...,l_e$. Each involves the patch merging module outputting $c_k^e$ patches of size $h_k \times w_k$ and $n_k^e$ VSSM-CA blocks processing these patches. At the end of the encoder, the output patches are fed into the convolution compression module, which uses a CNN to compress the patches to the designated number of channel uses. After that, the compressed patches are normalized for power and transmitted through the wireless channel. 

In this paper, we consider an end-to-end transmission through the Additive White Gaussian Noise (AWGN) channel or the Rayleigh fading channel. At the receiver, the MambaJSCC decoder first applies an equalizer, such as the minimum mean square error (MMSE) equalizer, to equalize the received signal $\mathbf{r}$ in the case of the Rayleigh fading channel. It then decodes the signal into $\mathbf{\hat{s}} \in \mathbb{R}^{3 \times H \times W}$ for reconstruction. The MambaJSCC decoder includes a CSI encoding module, $l_d$ decoding stages, and a convolution expanding module. Specifically, following the equalizer, the convolution expanding module first expands the equalized signal into $c^d_{l_d}$ 2D patches, each with size $h_{l_d} \times w_{l_d}$, to match the output patches from Stage $l_e$ in the encoder. These patches are then fed into $l_d$ concatenated stages, denoted as Stage $k,k=l_d,l_d-1,...,1$. For $k \ge 2$, each stage initially decodes the patches using $n_k^d$ VSSM-CA blocks along with the CSI vector $\mathbf{u_d}$, which is the output from the CSI encoding module in the decoder. The output patches are then divided into fewer but larger patches $c_{k-1}^d$ with size $h_{k-1} \times w_{k-1}$. This step is performed by a 2D patch division module that first applies layer normalization to the input patches, followed by a full-connection (FC) layer that expands the number of patches from $c^d_k$ to $c^d_{k-1}\times\frac{h_{k-1}w_{k-1}}{h_kw_k}$. The purpose of expanding the number of patches is to organize every $\frac{h_{k-1}w_{k-1}}{h_kw_k}$ patches into $c_{k-1}^d$ groups. Within each group, a pixel shuffle operation is then applied to cyclically select elements, thereby creating a new patch that is with $\frac{h_{k-1}w_{k-1}}{h_kw_k}$ times larger size. As a result, the module outputs $c^d_{k-1}$ new patches, each with a larger size of $h_{k-1}\times w_{k-1}$. In Stage 1, the 2D patch division module divides the output into three patches, each with size $H \times W$ for reconstruction.

The MambaJSCC encoder and decoder are jointly trained with the goal of minimizing the distortion between the input image and the reconstructed image. Therefore, the loss function $L$ can be derived as follows:
\begin{align}
  L=d(\mathbf{s},\mathbf{\hat{}s}),
\end{align}
where $d(\cdot)$ represents the distortion function.

\section{Structure of VSSM-CA Block}
\subsection{The VSSM Module}
\begin{figure*}[t]
  \begin{center}
    \includegraphics[scale=0.55]{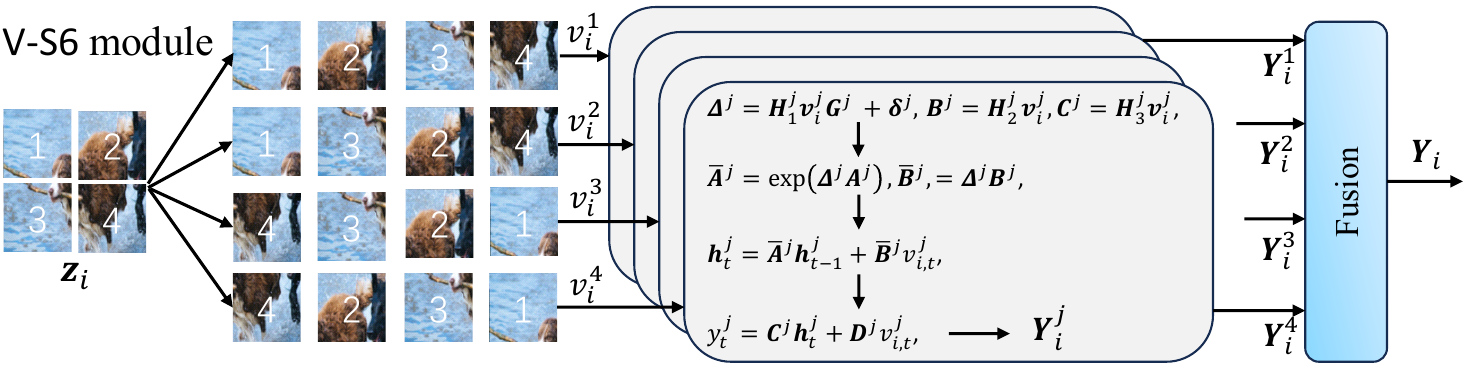}
  \end{center}
    \caption{{The computational flow of V-S6 module.}}
    \label{V-S6}
    \vspace{-0.3 cm}
\end{figure*}
The overall architecture of the VSSM-CA block is shown in Fig.\ref{MambaJSCC}(b). The block consists of two main components: the VSSM module and the CSI embedding module. For the VSSM module, it first normalizes the input patches using the layer-norm method and then feeds them into two branches. In the first branch, depicted in Fig.\ref{MambaJSCC}(b), an FC layer, a convolution layer and a SiLU activation function process the patches into higher-dimensional patches. Each of the $i$-th patch is denoted as $\mathbf{z}_i \in \mathbb{R}^{h_z\times w_z}, i=1,2,...,c_z$, where $c_z$ is the number of patches. These $c_z$ patches are then fed into $c_z$ V-S6 modules. 

The computational flow of the V-S6 module is described in Fig. \ref{V-S6}. Each vector $\mathbf{z}_i$ is flattened into four 1D vectors $\mathbf{v}_i^j \in \mathbb{R}^{D\times 1},j=1,2,3,4$ as follows:
\begin{alignat}{2}\label{expandz}
  &\mathbf{v}_i^1=\ vec(\mathbf{z}_i),
  &&\mathbf{v}_i^2=\ vec(\mathbf{z}_i^T),\nonumber\\
  &\mathbf{v}_i^3=\ R(vec(\mathbf{z}_i)),~~
  &&\mathbf{v}_i^4=\ R(vec(\mathbf{z}_i^T)),
\end{alignat}
where we use $D=h_z \times w_z$. The operation $vec(\cdot)$ denotes the vectorization of a matrix. The operation $R(\cdot)$ is the reversal of the order of elements in a vector or a matrix.

The four vectors are fed into four SSM modules, enabling the subsequent V-S6 module to acquire global information. For the $j$-th vector $\mathbf{v}_i^j$, the $j$-th SSM module utilizes seven learnable matrices $\mathbf{G}^j\in\mathbb{R}^{1 \times N}$, $\bm{\delta}^j\in\mathbb{R}^{N \times N}$ $\mathbf{H}_1^j$, $\mathbf{H}_2^j$, $\mathbf{H}_3^j\in\mathbb{R}^{N\times D}$, $\mathbf{A}^j \in \mathbb{R}^{N\times N}$ and $\mathbf{D}^j\in\mathbb{R}^{1\times 1}$ to model the state space $\mathbf{h}_t^j\in\mathbb{R}^{N\times 1}$. The module first computes the matrices $\mathbf{\Delta}^j$, $\mathbf{B}^j$ and $\mathbf{C}^j$ as follows:
\begin{equation}
  \mathbf{\Delta}^j=\mathbf{H}_1^j \mathbf{v}_i^j\mathbf{G}^j+\bm{\delta}^j,\ \mathbf{B}^j=\mathbf{H}_2^j \mathbf{v}_i^j,\ \mathbf{C}^j=(\mathbf{H}_3^j \mathbf{v}_i^j)^T,
\end{equation}
and then following the ZOH rule and the first-order Taylor series, the matrices $\mathbf{\bar{A}}^j$ and $\mathbf{\bar{B}}^j$ are computed as follows:
\begin{equation}
  \mathbf{\bar{A}}^j=\exp(\mathbf{\Delta}^j \mathbf{A}^j),\ \mathbf{\bar{B}}^j=\mathbf{\Delta}^j\mathbf{B}^j.
\end{equation}

Subsequently, the module iteratively computes the output for each element. For the $t$-th element of $\mathbf{v}_i^j$, the outputs are:
\begin{align}
  \mathbf{h}_t^j=\mathbf{\bar{A}}^j\mathbf{h}_{t-1}^j+\mathbf{\bar{B}}^j{v}_{i,t}^j, ~~  {y}_{i,t}^j=\mathbf{C}^j\mathbf{h}_t^j+\mathbf{D}^j{v}_{i,t}^j.
\end{align}
After the iterations, all the $y_{i,t}^j$ in this module are combined into a 2D matric $\mathbf{Y}_i^j\in\mathbb{R}^{h_z\times w_z}$. Then, we merge the outputs as follows:
\begin{align}
  \mathbf{Y}_i=\mathbf{Y}_i^1+(\mathbf{Y}_i^2)^T+R(\mathbf{Y}_i^3+(\mathbf{Y}_i^4)^T).
\end{align}
Therefore, the $c_z$ V-S6 modules output $c_z$ matrices. The algorithm of V-S6 module is summarized in Algorithm \ref{V-S6algorithm}.

In parallel, the normalized input patches are processed by an FC layer and a SiLU activation function in the second branch. The output of this branch is multiplied in element-wise with the output of the first branch, and then the result is projected back to the original dimension using another FC layer. A residual connection adds this result to the original input, yielding the final output.
\addtolength{\topmargin}{0.05in}
\begin{algorithm}[t]
  \hspace*{0.02in} {\bf \small{Input:}}
	\small{Input matrix $\mathbf{z}_i$;} \\
	\hspace*{0.02in} {\bf \small{Output:}}
	\small{Output matrix $\mathbf{Y}_i$;} 
	\caption{The computational flow of the $i$-th V-S6 module}
	\label{V-S6algorithm}
	\begin{algorithmic}[1]
    \STATE Expand $\mathbf{z_i}$ to $\mathbf{v}_i^1,\mathbf{v}_i^2,\mathbf{v}_i^3,\mathbf{v}_i^4$ according to (\ref{expandz});
    \FOR {$j=1$ to 4}
    \STATE $\mathbf{\Delta}^j=\mathbf{H}_1^j \mathbf{v}_i^j\mathbf{G}^j+\bm{\delta}^j,\ \mathbf{B}^j=\mathbf{H}_2^j \mathbf{v}_i^j,\ \mathbf{C}^j=(\mathbf{H}_3^j \mathbf{v}_i^j)^T$;
    \STATE $\mathbf{\bar{A}}^j=\exp(\mathbf{\Delta}^j \mathbf{A}^j),\ \mathbf{\bar{B}}^j=\mathbf{\Delta}^j\mathbf{B}^j$;
    \STATE $\mathbf{h}_0^j=\mathbf{0}$;
    \FOR {$t=1$ to $N$}
    \STATE $\mathbf{h}_t^j=\mathbf{\bar{A}}^j\mathbf{h}_{t-1}^j+\mathbf{\bar{B}}^j{v}_{i,t}^j$;
    \STATE ${y}_{i,t}^j=\mathbf{C}^j\mathbf{h}_t^j+\mathbf{D}^j{v}_{i,t}^j$;
    \ENDFOR
    \STATE Forming matrix $\mathbf{Y}_i^j$ using all $y_{i,t}^j$;
    \ENDFOR
    \STATE $\mathbf{Y}_i=\mathbf{Y}_i^1+(\mathbf{Y}_i^2)^T+R(\mathbf{Y}_i^3+(\mathbf{Y}_i^4)^T)$.
	\end{algorithmic}
  
\end{algorithm}
\subsection{The CSI Embedding Method}
The CSI embedding method, including CSI encoding modules and CSI embedding modules, is responsible for embedding CSI into the VSSM module, enhancing the single model's ability to handle various channel conditions.

In the MambaJSCC, the CSI encoding module includes sine position coding, an FC layer, a Swish activation function, and another FC layer. The module is responsible for coding the CSI into a vector and injecting it into the VSSM-CA blocks. All VSSM-CA blocks in the encoder share a single CSI encoding module, and similarly, a distinct CSI encoding module is shared by all VSSM-CA blocks within the decoder. Injecting the same CSI vector into each VSSM-CA block ensures that each stage of the model obtains accurate SNR information, thereby avoiding information loss that could occur with increased depth. This approach also eliminates the need to propagate SNR information to deeper layers, allowing the model to effectively encode the source information based on SNR, which conserves model capacity. Encoding the CSI into a higher-dimensional vector enhances the model's ability to utilize it. 

The CSI embedding module within the VSSM-CA block takes the CSI vector as input and expands its length to match the number of patches using an FC layer. As a result, each element of the expanded CSI vector can be added to each patch. This addition operation is effective because it is analogous to concatenating two features and processing them with a shared model. The design also reduces delay and model size. The FC layer also allows different VSSM-CA blocks to flexibly utilize the CSI. 

\section{EXPERIMENTS RESULTS}
\subsection{Experiment Setup}
In the simulation, we use the DIV2K dataset, which contains 800 varied 2K resolution images from a wide range of real-world scenes for training, along with an additional 100 images for testing. During raining, images are randomly cropped into $256\times 256$ patches.

The performance of the proposed MambaJSCC is evaluated using peak-signal-to-noise ratio (PSNR). Computational complexity is assessed by the multiply-accumulate operations (MACs), which count the number of multiplications and additions required for inference. We also measure the inference delay for a comprehensive evaluation. The model size is quantified by the number of parameters. Both MACs and parameters size are calculated by the Pytorch-OpCounter library, and the ID is measured on the NVIDIA A40 GPU.

We conduct a comparative analysis between MambaJSCC and SwinJSCC \cite{SwinJSCC}, with main comparison schemes as follows: 1) \textbf{MambaJSCC w/o CA}: MambaJSCC without the CSI encoding module and CSI embedding module; 2) \textbf{SwinJSCC w/o SA\&RA}: SwinJSCC without the Channel ModNet and Rate ModNet; 3) \textbf{MambaJSCC with CSI embedding}: MambaJSCC with CSI embedding; 4) \textbf{SwinJSCC w/ SA}: SwinJSCC with the Channel ModNet, which is an attention-based module for channel adaptation; 5) \textbf{MambaJSCC with Channel ModNet}: this scheme uses the Channel ModNet instead of CSI embedding in MambaJSCC.

\begin{figure}[t]
  \begin{center}
    \includegraphics[scale=0.6]{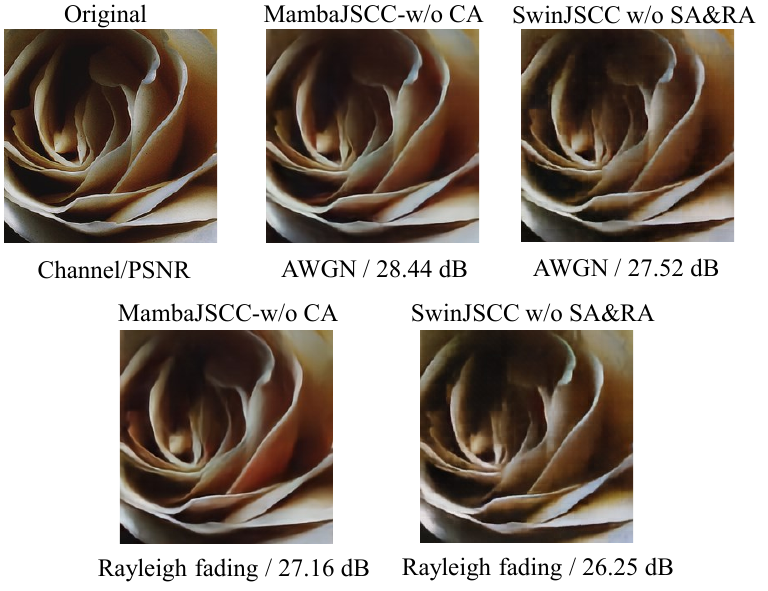}
  \end{center}
    \caption{{Visual comparison between MambaJSCC w/o CA and SwinJSCC w/o SA\&RA with $N_m=N_s=6$ under the AWGN and the Rayleigh fading channel with SNR=$5$ dB.}}
    \label{visual_results}
    \vspace{-0.5 cm}
\end{figure}
For the basic configurations of MambaJSCC, we adopt four stages in both the encoder and decoder, with $[n_1^e,n_2^e,n_3^e,n_4^e]=[n_1^d,n_2^d,n_3^d,n_4^d]=[2,2,6,2]$ and $[c_1^e,c_2^e,c_3^e,c_4^e]=[c_1^d,c_2^d,c_3^d,c_4^d]=[128,192,256,320]$. The CSI vector length is set to 128. To vary the model scales, we fix the block number in Stage 3  at $n_3^e=n_3^d$ and adjust their values $N_m$. Similarly, for SwinJSCC w/o SA\&RA and SwinJSCC w/ SA, we use the same method to modify the model size. The number of the Swin Transformer blocks in Stage 3 of the encoder and decoder for both models is denoted as $N_s$. The channel bandwidth ratio is set to $\frac{3}{128}$. We train all models using the Adam optimizer over 600 epochs on an NVIDIA A40 GPU, with an initial learning rate of $10^{-4}$ and a batch size of 4. The loss function is the mean square error (MSE) loss.

\subsection{Results of MambaJSCC w/o CA}
Fig. \ref{visual_results} visualizes the reconstruction results of MambaJSCC w/o CA and SwinJSCC w/o SA\&RA with $N_m=N_s=6$ over different channels. We can observe that in this example, the visual reconstruction quality of MambaJSCC w/o CA is significantly better than that of SwinJSCC w/o SA\&RA. For example, the edges and colors of the flower in the images reconstructed by MambaJSCC w/o CA more closely resemble the original image than those in images reconstructed by SwinJSCC w/o SA\&RA.

\begin{figure}[t]
  \begin{center}
    \includegraphics[scale=0.385]{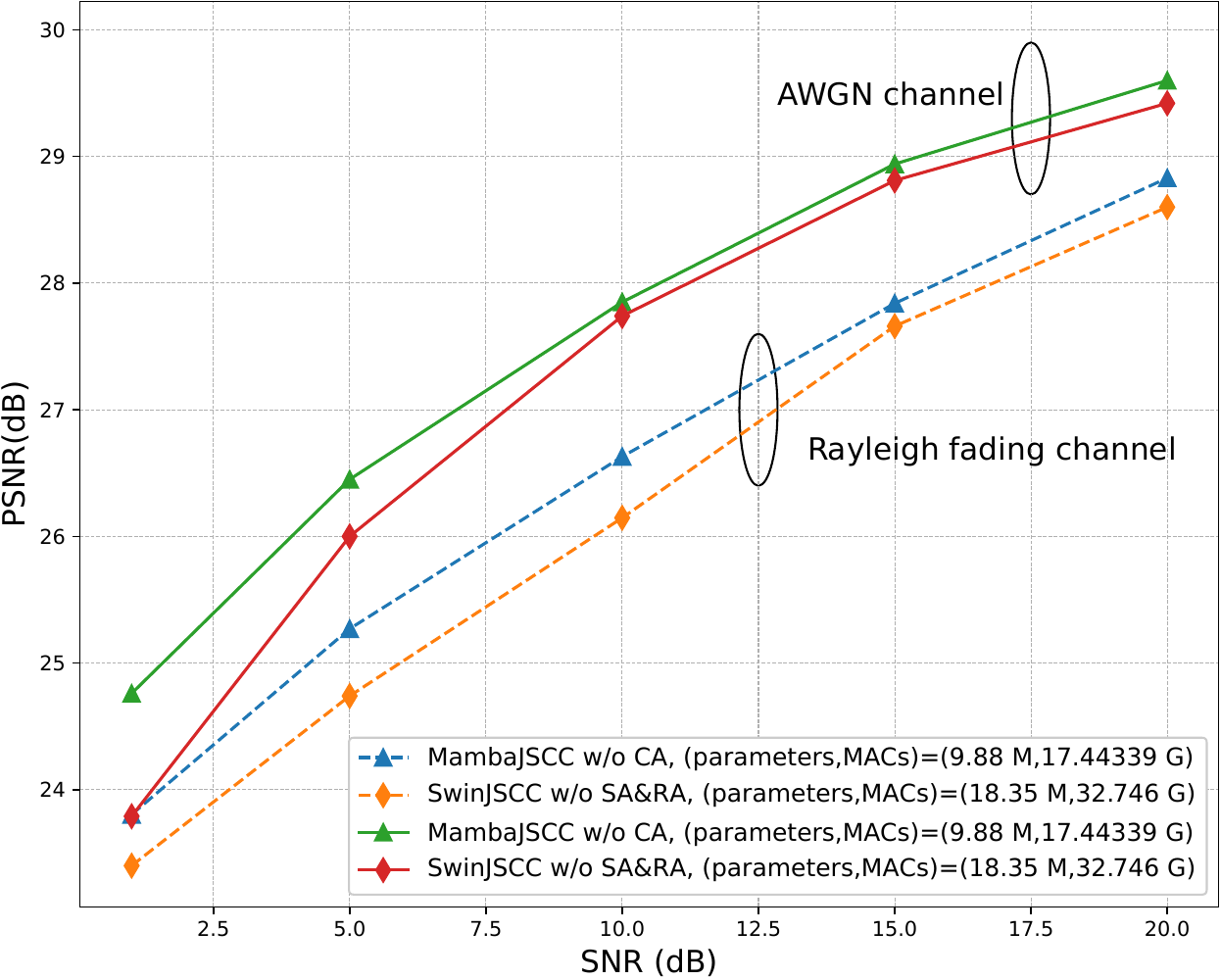}
  \end{center}
    \caption{{The PSNR performance of MambaJSCC w/o CA and SwinJSCC w/o SA\&RA versus SNR under the AWGN and the Rayleigh fading channels. $N_m$ and $N_s$ are set to $6$.}}
    \label{VSSM}
\end{figure}

Fig. \ref{VSSM} displays the PSNR performance of MambaJSCC w/o CA and SwinJSCC w/o SA\&RA versus SNR under both the AWGN and the Rayleigh fading channels. The legend provides the model size and computational complexity, represented by parameters and MACs, respectively. We can see that MambaJSCC w/o CA requires $15.30$ billion fewer MACs and $8.47$ million fewer parameters compared with SwinJSCC w/o SA\&RA. Meanwhile, we can see from Table I that the ID of MambaJSCC w/o CA is 39.12 ms shorter than that of SwinJSCC w/o SA\&RA. Even with significant reductions in both MACs and parameters, and a lower ID, the proposed MambaJSCC w/o CA still outperforms SwinJSCC w/o SA\&RA in terms of PSNR. For example, on the Rayleigh fading channel, Mamba JSCC w/o CA achieves a $0.53$ dB gain at an SNR of $5$ dB. For the AWGN channel, Mamba JSCC w/o CA achieves a $0.97$ dB gain at an SNR of $1$ dB.
\begin{table*}[t]
  \renewcommand\arraystretch{1.5}
  \centering
  \caption{{MACs, ID,  Parameters and PSNR at SNR=10 dB of models with different model sizes and computational complexity.}}
  \label{tab1}
  \begin{tabular}{|c|c|c|c|c|c|c|}\hline
  
  \multicolumn{2}{|c|}{Models}& MACs (G) & ID (ms) & Parameters (M) & PSNR, the AWGN channel (dB) & PSNR, the Rayleigh fading channel (dB)\\
  \hline 
  \multirow[vpos]{2}{*}{MambaJSCC} &$N_m=6$ & 17.44339 & 31.95 & 9.88 & 27.85 &26.63\\ \cline{3-7}  
  \multirow[vpos]{2}{*}{w/o CA} & $N_m=12$ &22.36960 & 44.59 & 14.68& 27.98 & 26.64\\\cline{3-7} 
  & $N_m=18$  & 27.29581 & 56.65 & 19.48 & 28.21 &26.79\\\hline 
  {SwinJSCC }& $N_s=2$ &26.27390 & 54.74 & 12.03 & 27.23&26.13\\\cline{3-7}  
  w/o SA\&RA & $N_s=6$ & 32.74564 & 71.07 & 18.35 & 27.74&26.15\\\hline
  \end{tabular}
  \vspace{-0.3 cm}
  \end{table*}

Furthermore, we investigate the potential performance gains of MambaJSCC w/o CA compared to the SwinJSCC w/o SA\&RA, considering similar model sizes and computational complexities. We adjust $N_m$ and $N_s$, with the results shown in Table \ref{tab1}. We first fix $N_s=6$ for SwinJSCC w/o SA\&RA, then increase $N_m$ for MambaJSCC w/o CA to $12$ and $18$ to expand the model size and computational complexity. The results show that when $N_m$ is increased to $18$, MambaJSCC w/o CA achieves a similar model size and computational complexity as SwinJSCC w/o SA\&RA, with slightly more parameters and fewer MACs. Under these conditions, the performance gain increases from $0.11$ dB to $0.47$ dB at an SNR of $10$ dB under the AWGN channel, and from $0.48$ dB to $0.64$ dB at an SNR of $10$ dB under the Rayleigh fading channel. Moreover, we also reduce $N_s$ for SwinJSCC w/o SA\&RA to $2$ to decrease its model size and computational complexity. It can be observed that SwinJSCC w/o SA\&RA with $N_s=2$ has a similar model size and computational complexity to MambaJSCC w/o CA with $N_m=12$, but with slightly more MACs and fewer parameters. In this configuration, the performance gain increases from $0.11$ dB to $0.75$ dB under the AWGN channel at an SNR of $10$ dB, and from $0.48$ dB to $0.51$ dB under the Rayleigh fading channel at the same SNR. Experimental results demonstrate that MambaJSCC w/o CA model can achieve significant performance gains, even with similar model size and computational complexity. 

\subsection{Results of MambaJSCC with CSI embedding}

\begin{figure}[t]
  \begin{center}
    \includegraphics[scale=0.385]{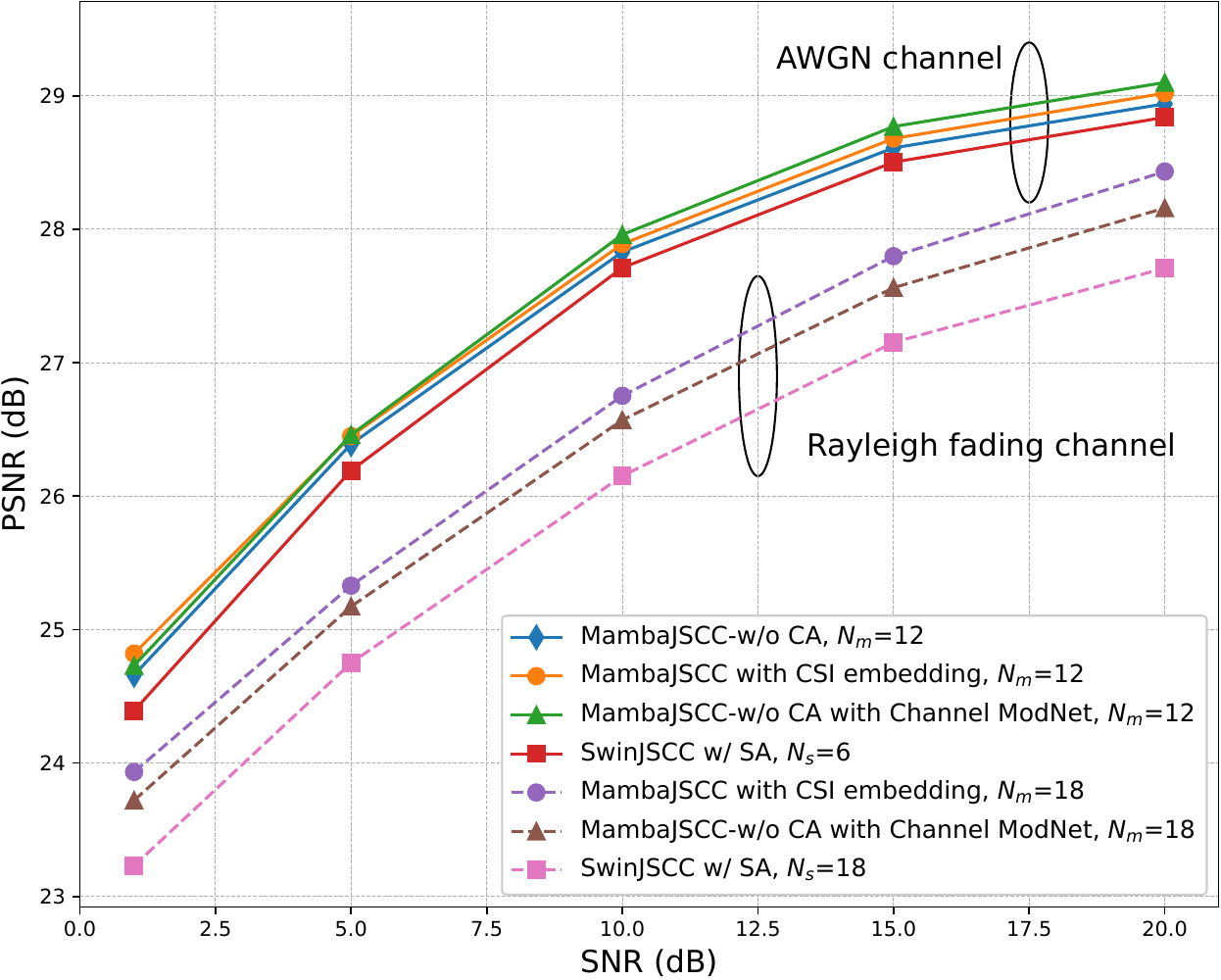}
  \end{center}
    \caption{{The PSNR performance with different channel adaptation methods.}}
    \label{CSI}
    \vspace{-0.3 cm}
\end{figure}

Fig. \ref{CSI} illustrates the performance of SwinJSCC w/ SA and MambaJSCC with different channel adaptation methods versus SNR under the AWGN and the Rayleigh fading channel. The $N_m$ is set to $12$ for the AWGN channel and $18$ for the Rayleigh fading channel. $N_s$ is set to $6$. We can see that MambaJSCC with CSI embedding significantly outperforms SwinJSCC w/ SA across all SNRs in the both channels. For example, in the AWGN channel, a maximum gain of $0.43$ dB is achieved at SNR=$1$ dB, and the minimum gain is $0.19$ dB at SNR=$20$ dB. For the Rayleigh fading channel, the performance improvements at different SNRs are consistent, with an average gain of $0.65$ dB. Moreover, the parameter size and the computational complexity of MambaJSCC with CSI embedding are significantly lower than those of SwinJSCC w/ SA. As shown in Table \ref{tab2}, compared to SwinJSCC w/ SA with $N_s=6$, MambaJSCC with CSI embedding requires $12.32$ billion fewer MACs, $11.70$ million fewer parameters for $N_m=12$, and $7.39$ billion fewer MACs and $6.90$ million fewer parameters for $N_m=18$. Furthermore, ID is reduced by $29.48$ ms and $17.30$ ms for $N_m=12$ and $N_m=18$ respectively.

Furthermore, we compare MambaJSCC with different channel adaptation modules. It can be observed that under the AWGN channel, MambaJSCC with CSI embedding and MambaJSCC w/o CA with Channel ModNet outperform MambaJSCC w/o CA, achieving similar performance. For example, MambaJSCC with CSI embedding achieves a $0.09$ dB higher PSNR at SNR=$1$ dB but performs $0.08$ dB lower as the SNR increases to $20$ dB compared with MambaJSCC w/o CA with Channel ModNet. For the Rayleigh fading channel, MambaJSCC with CSI embedding outperforms MambaJSCC w/o CA with Channel ModNet. The maximum gain is $0.28$ dB at SNR=$20$ dB, and at SNR=$5$ dB, the gain is $0.16$ dB.

Although the performance of the proposed CSI embedding is comparable to that of the Channel ModNet in MambaJSCC, the advantage of the CSI embedding method lies in its significantly low model size and computational complexity. As shown in Table \ref{tab2}, the CSI embedding method requires only $0.290~(0.41)$ million of extra MACs, $1.00~(1.12)$ ms for additional ID, and $0.297~(0.297)$ million extra parameters at $N_m=12~(18)$. In contrast, the Channel ModNet requires $872.13$ million extra MACs and $9.86$ million extra parameters for both configurations. These results indicate that the CSI embedding method significantly reduces the parameter size and computational complexity compared to the Channel ModNet, while maintaining comparable or superior performance. This efficiency makes the CSI embedding a promising method for channel adaptation in JSCC. 

\begin{table}[t]
  \renewcommand\arraystretch{1.5}
  \centering
  \caption{{MACs, ID and Parameters of different models with diverse channel adaptation methods.}}
  \label{tab2}
  \begin{tabular}{|p{1.4cm}<{\centering}|p{2.0cm}<{\centering}|p{0.95cm}<{\centering}|p{0.65cm}<{\centering}|p{1.5cm}<{\centering}|}\hline

  \multicolumn{2}{|c|}{Models}& MACs(G) & ID(ms)& Parameters(M) \\
  \hline
  \multirow[vpos]{3}{*}{MambaJSCC} & w/o CA & 22.36960 & 44.59 & 14.68 \\\cline{2-5}
  & CSI embedding & 22.36989 & 45.59 &  14.97\\\cline{2-5}
  \multirow[vpos]{1}{*}{\ $N_m=12$}&  w/o CA with  & \multirow[vpos]{2}{*}{23.24113} & \multirow[vpos]{2}{*}{49.79} &  \multirow[vpos]{2}{*}{24.54}  \\ 
  &Channel ModNet & & &    \\   \hline 
  \multirow[vpos]{3}{*}{MambaJSCC} & w/o CA & 27.29581 & 56.65 &  19.48\\\cline{2-5}
  & CSI embedding & 27.29622 & 57.77 & 19.77 \\\cline{2-5}
  \multirow[vpos]{1}{*}{\ $N_m=18$}& w/o CA with &  \multirow[vpos]{2}{*}{28.16734} & \multirow[vpos]{2}{*}{61.24} &  \multirow[vpos]{2}{*}{29.34}  \\
  &Channel ModNet & & &    \\   \hline 
  
  \multicolumn{2}{|c|}{SwinJSCC w/ SA, $N_s=6$}& 34.68512 & 75.07 &  26.67\\\hline
   
  \end{tabular}
  \vspace{-0.3 cm}
  \end{table}

\section{CONCLUSION}
In this paper, we propose MambaJSCC, a novel lightweight and efficient JSCC architecture for wireless image transmission. MambaJSCC utilizes VSSM for effective representation learning and employs a CSI embedding method, integrating the CSI vector from CSI encoding module into each VSSM-CA block to effectively handle various channel conditions. Experimental results confirm that the proposed MambaJSCC achieves remarkable performance across a range of channel conditions. Moreover, MambaJSCC requires fewer parameters, MACs and exhibits a low ID, effectively overcoming the challenges of high computational complexity and large model size associated with Transformer. Therefore, MambaJSCC presents a promising solution for semantic communication systems that demand real-time responsiveness, high energy efficiency, and operate within storage constraints.

\footnotesize
\bibliographystyle{IEEEtran}
\bibliography{reference}{}

\begin{thebibliography}{10}
\providecommand{\url}[1]{#1}
\csname url@samestyle\endcsname
\providecommand{\newblock}{\relax}
\providecommand{\bibinfo}[2]{#2}
\providecommand{\BIBentrySTDinterwordspacing}{\spaceskip=0pt\relax}
\providecommand{\BIBentryALTinterwordstretchfactor}{4}
\providecommand{\BIBentryALTinterwordspacing}{\spaceskip=\fontdimen2\font plus
\BIBentryALTinterwordstretchfactor\fontdimen3\font minus
  \fontdimen4\font\relax}
\providecommand{\BIBforeignlanguage}[2]{{%
\expandafter\ifx\csname l@#1\endcsname\relax
\typeout{** WARNING: IEEEtran.bst: No hyphenation pattern has been}%
\typeout{** loaded for the language `#1'. Using the pattern for}%
\typeout{** the default language instead.}%
\else
\language=\csname l@#1\endcsname
\fi
#2}}
\providecommand{\BIBdecl}{\relax}
\BIBdecl

\bibitem{Qin2}
D.~Gündüz, Z.~Qin, I.~E. Aguerri, H.~S. Dhillon, Z.~Yang, A.~Yener, K.~K.
  Wong, and C.-B. Chae, ``{Beyond Transmitting Bits: Context, Semantics, and
  Task-Oriented Communications},'' \emph{IEEE Journal on Selected Areas in
  Communications}, vol.~41, no.~1, pp. 5--41, 2023.

\bibitem{Qin1}
Z.~Qin, F.~Gao, B.~Lin, X.~Tao, G.~Liu, and C.~Pan, ``{A Generalized Semantic
  Communication System: From Sources to Channels},'' \emph{IEEE Wireless
  Communications}, vol.~30, no.~3, pp. 18--26, 2023.

\bibitem{shao}
J.~Liu, S.~Shao, W.~Zhang, and H.~V. Poor, ``{An Indirect Rate-Distortion
  Characterization for Semantic Sources: General Model and the Case of Gaussian
  Observation},'' \emph{IEEE Transactions on Communications}, vol.~70, no.~9,
  pp. 5946--5959, 2022.

\bibitem{wei}
J.~Xu, B.~Ai, N.~Wang, and W.~Chen, ``{Deep Joint Source-Channel Coding for CSI
  Feedback: An End-to-End Approach},'' \emph{IEEE Journal on Selected Areas in
  Communications}, vol.~41, no.~1, pp. 260--273, 2023.

\bibitem{CDDM}
T.~Wu, Z.~Chen, D.~He, L.~Qian, Y.~Xu, M.~Tao, and W.~Zhang, ``{CDDM: Channel
  Denoising Diffusion Models for Wireless Semantic Communications},''
  \emph{IEEE Transactions on Wireless Communications}, pp. 1--1, 2024.

\bibitem{text}
S.~Yao, K.~Niu, S.~Wang, and J.~Dai, ``{Semantic Coding for Text Transmission:
  An Iterative Design},'' \emph{IEEE Transactions on Cognitive Communications
  and Networking}, vol.~8, no.~4, pp. 1594--1603, 2022.

\bibitem{video}
S.~Wang, J.~Dai, Z.~Liang, K.~Niu, Z.~Si, C.~Dong, X.~Qin, and P.~Zhang,
  ``{Wireless Deep Video Semantic Transmission},'' \emph{IEEE Journal on
  Selected Areas in Communications}, vol.~41, no.~1, pp. 214--229, 2023.

\bibitem{gundu2019}
E.~Bourtsoulatze, D.~Burth~Kurka, and D.~Gündüz, ``{Deep Joint Source-Channel
  Coding for Wireless Image Transmission},'' \emph{IEEE Transactions on
  Cognitive Communications and Networking}, vol.~5, no.~3, pp. 567--579, 2019.

\bibitem{SwinJSCC}
K.~Yang, S.~Wang, J.~Dai, X.~Qin, K.~Niu, and P.~Zhang, ``{SwinJSCC: Taming
  Swin Transformer for Deep Joint Source-Channel Coding},'' \emph{arXiv
  preprint arXiv:2308.09361}, 2023.

\bibitem{Ze}
Z.~Liu, Y.~Lin, Y.~Cao, H.~Hu, Y.~Wei, Z.~Zhang, S.~Lin, and B.~Guo, ``{Swin
  Transformer: Hierarchical Vision Transformer using Shifted Windows},'' in
  \emph{Proc. IEEE/CVF ICCV}, 2021, pp. 9992--10\,002.

\bibitem{Xu}
J.~Xu, B.~Ai, W.~Chen, A.~Yang, P.~Sun, and M.~Rodrigues, ``{Wireless Image
  Transmission Using Deep Source Channel Coding With Attention Modules},''
  \emph{IEEE Transactions on Circuits and Systems for Video Technology},
  vol.~32, no.~4, pp. 2315--2328, 2022.

\bibitem{VSSM}
Y.~Liu, Y.~Tian, Y.~Zhao, H.~Yu, L.~Xie, Y.~Wang, Q.~Ye, and Y.~Liu, ``{Vmamba:
  Visual state space model},'' \emph{arXiv preprint arXiv:2401.10166}, 2024.

\bibitem{S6}
A.~Gu and T.~Dao, ``{Mamba: Linear-time sequence modeling with selective state
  spaces},'' \emph{arXiv preprint arXiv:2312.00752}, 2023.

\end{thebibliography}
\end{document}